\documentclass[fleqn,10pt]{wlscirep}

\usepackage{graphicx}
\usepackage{dcolumn}
\usepackage{bm}

\usepackage[utf8]{inputenc}
\usepackage[T1]{fontenc}
\usepackage{mathptmx}
\usepackage{physics}

\begin{document}


\title{Robust and fast post-processing of single-shot spin qubit detection events with a neural network} 

\author[1]{Tom Struck}%
\affil[1]{JARA-FIT Institute for Quantum Information, Forschungszentrum J\"ulich GmbH and RWTH Aachen University, Aachen, Germany}
\author[1]{Javed Lindner}%
\author[1]{Arne Hollmann}%
\author{Floyd Schauer}%
\affil[2]{Institut für Experimentelle und Angewandte Physik, Universit\"at Regensburg, Regensburg, Germany}
\author[2]{Andreas Schmidbauer}%
\author[2]{Dominique Bougeard}%
\author[1,*]{Lars R. Schreiber}%
\affil[*]{lars.schreiber@physik.rwth-aachen.de}


\date{\today}

\begin{abstract}
Establishing low-error and fast detection methods for qubit readout is crucial for efficient quantum error correction. Here, we test neural networks to classify a collection of single-shot spin detection events, which are the readout signal of our qubit measurements. This readout signal contains a stochastic peak, for which a Bayesian inference filter including Gaussian noise is theoretically optimal. Hence, we benchmark our neural networks trained by various strategies versus this latter algorithm. Training of the network with 10$^{6}$ experimentally recorded single-shot readout traces does not improve the post-processing performance. A network trained by synthetically generated measurement traces performs similar in terms of the detection error and the post-processing speed compared to the Bayesian inference filter. This neural network turns out to be more robust to fluctuations in the signal offset, length and delay as well as in the signal-to-noise ratio. Notably, we find an increase of 7\,\% in the visibility of the Rabi-oscillation when we employ a network trained by synthetic readout traces combined with measured signal noise of our setup. Our contribution thus represents an example of the beneficial role which software and hardware implementation of neural networks may play in scalable spin qubit processor architectures. 
\end{abstract}

\maketitle 
\section{Introduction}

Fast and high-fidelity single-shot readout of qubits is vital for the realisation of quantum information processing. Since quantum error correction schemes require frequent qubit readout \cite{Fowler2012}, the qubit measurement time should not be much longer than the qubit manipulation time to avoid speed limitations. The readout scheme depends upon the specific qubit realisation and can be discriminated into two categories \cite{DAnjou2014}: The measurement signal starts either immediately after the trigger of the detection process \cite{Robledo2011}, or it is delayed randomly by a turn-on time \cite{Elzerman2004}. While spin-to-charge conversion of a singlet-triplet spin readout by Pauli-spin blockade falls into the first category \cite{Petta2005,Maune2012}, single-spin detection by energy-dependent tunneling to a weakly tunnel-coupled reservoir falls into the second \cite{Elzerman2004,Morello2010}. For the latter, the analog measurement signal is often post-processed by peak-signal filters to assign a binary qubit readout. Examples for peak-signal filter are wavelet edge detection \cite{Prance2015}, signal threshold \cite{Elzerman2004,Nowack2011} and slope threshold after filtering the signal with total variation denoising \cite{Hollmann2020,Struck2020}.

If only one single spin detection cycle is considered, a Bayesian inference filter capturing the tunneling constants and typical noise is optimal \cite{DAnjou2014}. The readout speed with the Bayesian filter can be improved by adaptive decisions which allow to balance measurement time versus read-out fidelity \cite{DAnjou2016}. As the signal-to-noise ratio (SNR) of the detection signal is lowered for qubits in a dense array \cite{Li2017} or for charge detectors operating at elevated temperature, post-processing robust to low SNR is essential for future quantum computing architectures and hot electron spin qubits \cite{Vandersypen2017}, motivating the testing of alternatives to the theoretically optimal Bayesian method.

\begin{figure}
\includegraphics[width=\linewidth]{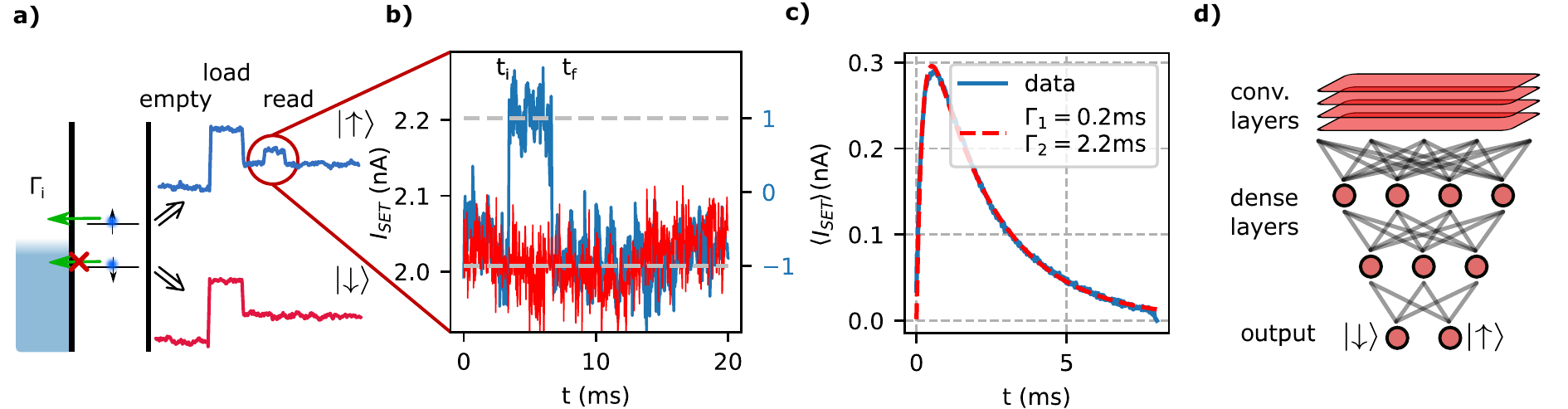}
\caption{Single-shot measurement traces and architecture of the neural network. a) Sketch of the Zeeman energy-resolved tunneling from the QD to the reservoir with a tunnel rate $\Gamma_i$. Two typical SET current traces showing the cycle of emptying and loading stage of the QD followed by the qubit readout stage are also shown. Only when the qubit state is $\ket{\uparrow}$-spin, the electron tunnels out of the QD at time $t_i$ and a $\ket{\downarrow}$-spin tunnels on the QD at time $t_f$ resulting in a peak in the SET current (blue trace). b) Zoom-in of two measured SET current traces spanning the readout stage. The blue and red traces are classified to be $\ket{\uparrow}$ and $\ket{\downarrow}$, respectively. c) Current signal spanning the readout stage averaged for many $\ket{\downarrow}$ and $\ket{\uparrow}$ traces (blue), fitted by equation \ref{eq:1} (dashed red line). d) Architecture of the network used to classify peak traces. One trace spanning the readout time is used as input (top) and classification is output by two neurons (bottom), giving the probability of $\ket{\downarrow}$ and $\ket{\uparrow}$ trace.}
\label{fig:1}
\end{figure}

\section{Results}

Here, we report on the performance of neural networks, which have been previously used to tune the electrostatics of devices\cite{Lennon,Kalantre,Nguyen,Zwolak}, to post-process single spin readout by spin-to-charge conversion. We compare their robustness and post-processing time to a Bayesian inference filter. We find that a neural network can perform similarly to the Bayesian inference filter on synthetic data and slightly outperforms it on real data, if it is made more robust by variations in the training data. Considerably better performance is achieved on the classification of measured data, when the neural network is trained with synthetic traces combined with measured noise.

Our qubit system is a single electron spin qubit trapped in an electrostatically defined Si/SiGe quantum dot (QD) \cite{Hollmann2020}. The qubit is encoded by the electron spin states $\ket{\uparrow}$ and $\ket{\downarrow}$, which are energetically split by an in-plane magnetic field of 668\,mT. Our detection signal consists of the single-shot readout of spin orientations via spin-to-charge conversion: Setting the chemical potential of the two-dimensional electron reservoir (electron temperature 114\,mK) as plotted in Fig. \ref{fig:1}a at time $t=0$\,s, energetically only an electron in the $\ket{\uparrow}$ state can tunnel into the reservoir after a time $t_i$ following Poisson statistics \cite{Elzerman2004}. At a time $t_f$, the empty QD is reinitialized by an electron in a $\ket{\downarrow}$-state from the reservoir. The two tunnel events are detected by the current $I_{SET}$ of a capacitively coupled single-electron transistor (SET). Signal traces for $\ket{\uparrow}$ and $\ket{\downarrow}$ events are shown exemplarily in Fig. \ref{fig:1}b. Averaging the $I_{SET}$ traces of $3\cdot 10^5$  $\ket{\uparrow}$-events (Fig. \ref{fig:1}c ), we fit the distribution by

\begin{equation}
    \langle I_{SET} (t) \rangle \propto e^{-t\Gamma_f}(1-e^{-t\Gamma_i})
    \label{eq:1}
\end{equation}

\noindent where $I_{SET}(t)$ is the current signal, meaning that $\langle I_{SET}(t) \rangle$ is proportional to the probability that the QD is unoccupied (Fig. \ref{fig:1}c). $\Gamma_{i}$ and $\Gamma_{f}$ are the tunneling rates out and into the QD, respectively. For the plotted example, we find $\Gamma_i^{-1}=0.20\pm 0.07\,$ms and $\Gamma_f^{-1}=2.20\pm0.004\,$ms. 

We consider a network implemented in tensorflow \cite{tensorflow2015-whitepaper} with keras and investigate its post-processing performance to classify the $I_{SET}$ traces into $\ket{\uparrow}$- and $\ket{\downarrow}$-events. The input layer is connected to four 1D convolutional layers with kernel sizes (101,51,25,10), a filter depth of (32,16,16,8) and ReLU \cite{Nair2010} activation (Fig. \ref{fig:1}d). Inbetween each convolutional layer, a maxpooling layer with size 3 is inserted. The convolutional layers feed into a dense network with three layers of size (64,32,2). The first two layers use ReLU  activation and the last layer uses softmax, since the last two neurons categorize into 1 and 0 for a $\ket{\uparrow}$- and $\ket{\downarrow}$-event, respectively. This neural network architecture was selected after testing variations of the network, both by changing it by hand and with Bayesian optimisation of some of the network parameters: We optimised the kernel sizes, filter depths of the convolutional layers, as well as the number and size of the dense layers and the dropout rate after each dense layer using the Bayesian optimisation.  
We find that sufficiently large networks have a similar error rate. If the network is too small, e.g. three small dense layers and no convolutional layers, the achieved accuracy decreases by approx. 4-5\%. A network without convolutional layers is also able to reach the same accuracy, but converges slower. Note, that too large networks become inefficient as far as training and evaluation time is concerned. The size of the architecture chosen here, represents the best compromise we found.

\begin{figure}
    \centering
    \includegraphics[width=\linewidth]{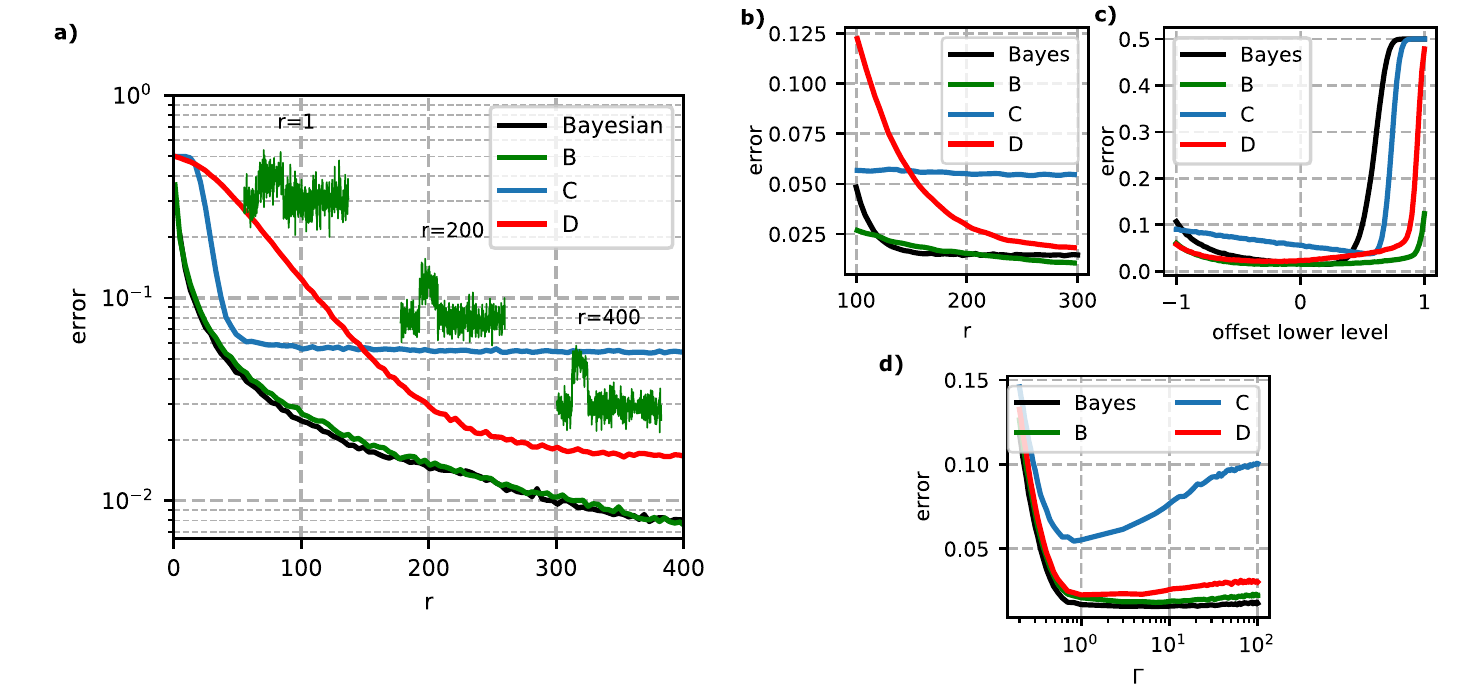}
    \caption{a) Classification error of the neural network $B$, trained with synthetic $\ket{\uparrow}$- and $\ket{\downarrow}$-traces, compared to classification by the Bayesian inference filter, considering Gaussian noise, $\Gamma = 1$ and various $r$. The network $C$ is trained on meausured data and network $D$ is trained with synthetic traces superimposed with typical noise from the experimental setup. The green inset traces are example input traces for $\ket{\uparrow}$-events at a given $r$.b)-d) Robustness of the Bayesian inference filter and the neural networks $B$, $C$ and $D$ to three parameters of synthesized traces superimposed by Gaussian noise. b) Robustness to the SNR $r$ variations. c) Robustness to an offset of the low-current level of $I_{SET}$. The x-axis here indicates the deviation of the lower level from the value of -1. d) Robustness to variations in the tunnel rates $\Gamma$.}
    \label{fig:net_vs_est}
\end{figure}

We train this network architecture by three qualitatively different sets of traces and will call theses trained networks $B$, $C$ and $D$. In all cases, we employ the neural network architecture explained above with the adam optimiser \cite{kingma2017} and categorical cross-entropy as the loss function. For the training of network $B$, we synthesize $\ket{\uparrow}$- and $\ket{\downarrow}$-traces with Gaussian noise and $\Gamma = \Gamma_f / \Gamma_i = 1$. We express the SNR by $r$, where $r$ is the power signal to-noise-ratio integrated over the average high signal time $\langle t_f-t_i\rangle$ \cite{DAnjou2014}
\begin{equation}
    r^{-1}=\int_0^{\langle t_f-t_i\rangle}dt\int_0^{\langle t_f-t_i\rangle}dt'\langle\delta\Psi(t)\delta \Psi(t')\rangle
\end{equation}
where $\Psi$ is the signal trace scaled between $-1$ and 1, $\delta\Psi(t)=\Psi(t)-\langle\Psi(t)\rangle$, the deviation of the signal from the noiseless signal. Examples of traces with three different $r$ are plotted as inset in Fig. \ref{fig:net_vs_est}a. The synthetic traces are generated such that the position of the lower level in the noiseless signal is at -1 and the high level is at 1. Then we add Gaussian noise to this trace. The network $B$ is trained by synthetic traces with various $r$. It is trained on $5\cdot 10^5$ traces with an equal distribution of traces synthesized with $r$ ranging from 1 to 400. Since we synthesize the training data, the correct labeling of the event traces is ensured. This is in contrast to the network $C$, which we train with 10$^{6}$ measured traces collected over two months of continuously running experiments with one device. During the measurements we tried to keep $\Gamma_i$ and $\Gamma_f$ in the range specified in Fig. \ref{fig:1}c. The measured data has to be classified, since we need labels for the training. We used a Bayesian inference filter for this classification \cite{DAnjou2014}. The network $D$ is trained by traces, which are synthesized similar to the traces for the network $B$, but instead of Gaussian noise we generate noise from the measured power spectrum of the experimental setup, which encompasses the qubit device and all the setup electronics representing our common noise sources. Since we synthesize the peak of these training traces as we do for the training data of the network $B$, labeling is $100\, \%$ correct, while the labeling of the training data of network $C$ is defective due to the error of the Bayesian inference filter.  

Next, we compare the classification error of the neural networks $B$, $C$ and $D$ to the Bayesian inference filter from Ref. \cite{DAnjou2014} (see equation C4 in the appendix therein) as a function of the SNR. In order to determine the classification error, defined by the sum of the $\ket{\downarrow}$-states labeled $\ket{\uparrow}$, we synthesized 100.000 $\ket{\uparrow}$- and 100.000 $\ket{\downarrow}$-traces with Gaussian noise, $\Gamma = 1$ and various $r$ values. The classification error of the network $B$ is nearly the same, compared to the Bayesian inference algorithm, within numerical uncertainties (Fig. \ref{fig:net_vs_est}a). This result is not surprising, since neural networks having a large enough size can emulate any function \cite{Leshno1993}. Remarkably, however, the network $B$ classifies synthetic traces of various $r$, as it has been trained by various $r$ values, meaning that the network $B$ is made robust against $r$ fluctuations.
The neural networks $C$ and $D$ show a significant larger classification error than the network $B$ and the Bayesian estimate. This is due to the fact that these networks are trained with at least partially experimental contribution, which are not captured by the synthetic traces used to determine the classification error in Fig. \ref{fig:net_vs_est}a. Specifically, the typical experimental noise is more involved than just Gaussian noise and the networks $C$ and $D$ were only trained by the measured SNR range. Note that the network $C$ also contains the classification error of the Bayesian inference filter used for labeling the training data set. The network $D$ is trained on synthetic traces superimposed with experimental noise. If we determine the classification error by such synthetic traces superimposed with the experimentally measured noise spectrum, the $D$ network achieves already a lower classification error (2.7 \%) compared to the Bayesian inference filter (5.7 \%).

In experiments typically a SNR of $r \approx 400$ is achievable \cite{Simmons2011,Yoneda2018}. According to Fig. \ref{fig:net_vs_est}a, we can achieve a low qubit detection error rate of less than $1 \%$ with the network $B$ as well as with the Bayesian inference filter. However, this result only refers to the ideal synthetic traces. Real signal traces contain noise with complicated noise spectral densities originating from e.g. interference noise, charge noise from an ensemble two-level fluctuators, and Johnson noise. Thermal excitation, spin relaxation or co-tunneling faster than the time-scale of $t_i$ and $t_f$ lead to defective signal traces as well. 
These latter effects can be suppressed by proper tuning of the ratio of the tunnel rates to measurement bandwidth and the ratio of the Zeeman energy to the electron temperature. Challenging are slow variations of the SNR $r$, the current offset of the SET and variations of the ratio between tunnel-in and -out rate $\Gamma$, due to low-frequency charge-noise or uncompensated cross-capacitive couplings to other gates. Therefore, we investigate the robustness of the post-processing filters with respect to these parameters in the following paragraph.

Fig. \ref{fig:net_vs_est}b shows the robustness of the different approaches to variations in the $r$-parameter. Since the networks do not get $r$ as an input parameter, the dependencies here are the same as in Fig. \ref{fig:net_vs_est} a. In contrast to Fig. \ref{fig:net_vs_est}a , now the Bayesian filter has a fixed $r$-parameter set to 200. Comparing the Bayesian inference filter to the network $B$, remarkably, the Bayesian inference filter performs generally worse than the neural network $B$, e.g. it shows up to 2\% additional error at $r=100$, and it deviates sharply from the optimum as the SNR decreases. Here the Bayesian algorithm seems to readily interpret single spikes in the noisy signal trace as peaks and labels them as $\ket{\uparrow}$ by error. Note that the error rate of the Bayesian inference filter saturates at an error above 1 \% for an increasing SNR. This at first sight surprising observation is caused by $\ket{\uparrow}$-traces classified as $\ket{\downarrow}$, since the peak is mistaken to be Gaussian noise as the amplitude of Gaussian noise is much lower than the Bayesian filter expected by the implemented $r=200$. Very important is the position of the overall $I_{SET}$ level: The Bayesian estimate quickly fails at predicting the correct result (Fig. \ref{fig:net_vs_est}c ), while the neural network $B$ shows no increase of error in a wide offset range of -0.75 to 0.75. Here, the offset is to be understood as the error in assigning -1 to the lower signal level, while the total amplitude is kept at 2. Finally, we investigate the robustness of the different methods with respect to the $\Gamma$ parameter (Fig. \ref{fig:net_vs_est}d). We observe that all post-processing approaches except for network $C$ remain accurate for $\Gamma>1$. If $\Gamma \ll 1$ and thus $\Gamma_f \ll \Gamma_i$, the classification error rises for every method, since the peak beginning starts to exceed the measurement window, i.e. the predefined length of the signal trace.

\begin{figure}
    \centering
    \includegraphics[width=\linewidth/2]{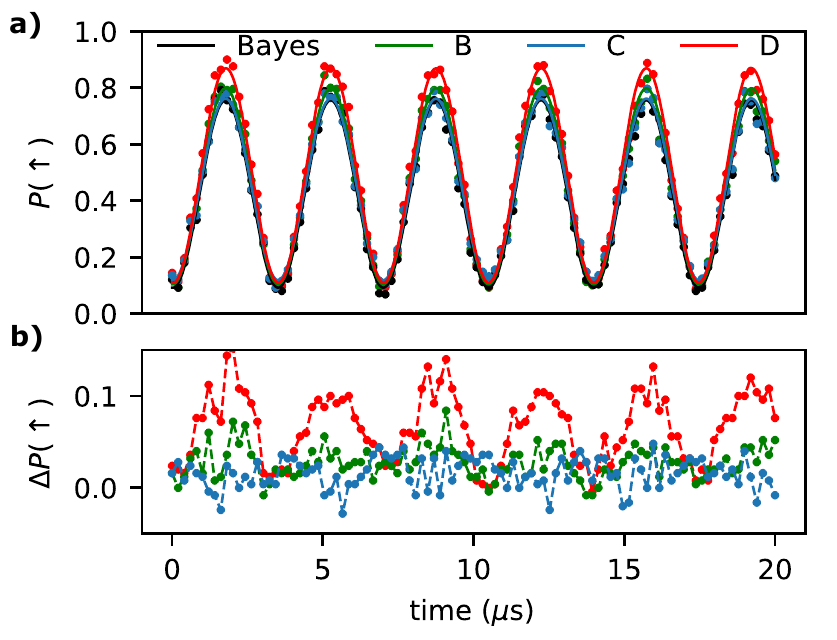}
    \caption{Detection of a Rabi-driven single electron spin analysed by different post-processing methods. a) Spin $\ket{\uparrow}$-probability $P(\ket{\uparrow})$ as a function of the driving time $t$. The same measurement traces are classified by the Bayesian inference filter, and the neural networks  $B$, $C$ and $D$. Solid lines are fits to Eq. \ref{eq:rabi}. b) $P(\ket{\uparrow})$ from panel a) classified by the neural networks $B$, $C$ and $D$ each subtracted by $P(\ket{\uparrow})$ classified by the Bayesian inference filter (colors equal to panel a).}   
    \label{fig:real_data_comp}
\end{figure}

We now apply the neural network approach and the Bayesian inference filter to classify experimentally recorded datasets. In contrast to synthetic traces, here, we face the fundamental problem that we cannot know a priori whether traces correspond to a $\ket{\uparrow}$- or $\ket{\downarrow}$-state. To benchmark our post-processing methods, we use the detection of Rabi-driven qubits following the formula 
\begin{equation}
P(\ket{\uparrow})=\frac{V}{2} \exp(-\frac{t}{T_R}) \cos(2\pi \nu t) + P_0,
\label{eq:rabi}
\end{equation}

\noindent where $t$ is the Rabi driving time, $\nu$ the Rabi-frequency, $T_R$ a Rabi-specific spin decay. $V$ and $P_0$ are the visibility and the offset of the Rabi oscillations, respectively. Thus, we expect a continuous variation of probabilities to find the $\ket{\uparrow}$ state ($P(\ket{\uparrow})$). Although $V$ and $P_0$ are reduced due to initialisation errors of the qubit and manipulation errors during Rabi driving (e.g off-resonant driving), it is reasonable to assume that the classification error of post-processing the readout traces reduces $V$ as well. Hence, we presume that a larger $V$ corresponds to a lower classification error if the same set of readout data is post-processed.
Before the data is analysed with the methods described above, it is rescaled and the lower level offset is removed. For rescaling we use the known height of the peak of $200\,$pA and the lower level is estimated with the median of the trace. Each data point in Fig. \ref{fig:real_data_comp}a is the average of 250 traces classified to be either in $\ket{\downarrow}$ or $\ket{\uparrow}$ state.
After fitting the data by Eq. \ref{eq:rabi} as shown in Fig. \ref{fig:real_data_comp}a, we find a visibility of $V=0.664\pm0.004$ for the analysis using the Bayesian inference filter. The network $C$, which is trained on real data, has a lower visibility of $V=0.644\pm0.004$ and thus does not perform better than the Bayesian inference filter. The neural network $B$, for which we find $V=0.695\pm0.004$, slightly outperforms the Bayesian estimation mainly due to the superior robustness of the neural network to variations in the $I_{SET}$ offset (Fig. \ref{fig:net_vs_est}c). The error in the lower level estimation can occur if a large portion of the signal is on the higher level, since in this case the median will estimate a wrong lower level current. The network $D$ reveals a significant larger visibility of $V=0.760\pm0.004$ compared to the Bayesian inference filter and all the other neural networks, thus its classification error is the lowest. Mainly the classification error of $\ket{\uparrow}$ traces is reduced (Fig. \ref{fig:real_data_comp}b). In contrast to network $B$ and the Bayesian inference filter, network $D$ is trained on the realistic noise spectrum, but does not suffer from the labeling problem of the real training data used for network $C$. Hence, this hybrid training approach outperforms all other training methods as well as the Bayesian inference filter based on a reasonably simple noise model.

Apart from the classification error and robustness to variations in a real experiment, the time $T$ required for the post-processing per trace is an important performance parameter. It adds up to the measurement time and can become critical for real-time feedback e.g. required during quantum error correction. In order to compare $T$ of the different classification traces, we let the whole post-processing run as efficiently as possible on the same computer equipped with an Intel i9 9900K processor. The differential equations for the Bayesian algorithm are solved using a Runge-Kutta method, in a Python script using numba just-in-time compilation. The neural networks runs with the tensorflow package. We find that $T$ for all neural networks is $\approx 50\, \mu$s and $\approx 200\, \mu$s for the Bayesian inference filter. Importantly, $T$ is of the same order of magnitude as the fastest reported experimental measurement times \cite{Vink2007}, hence representing a relevant contribution to classification processes if it runs on a computer. Note that a peak finder algorithm used in Refs. \cite{Hollmann2020, Struck2020} required approximately 100 times longer. $T$ of the Bayesian inference filter might be boosted by hardware encoding of the algorithm e.g. in field programmable gate arrays. This is also possible with the neural network in dedicated neural network hardware chips. Hence, both methods present advantages for low-temperature and low-power control electronics in the future.   

\section{Discussion}
In summary, we have shown that the neural network approach is a competitive alternative to post-processing of single-shot spin detection events by a Bayesian inference filter. The processing speeds are similar, with a slight advantage for the neural network.
We have benchmarked the performance of the neural network versus a Bayesian inference filter on synthetic and experimental data, using different training methods for the network. Since the Bayesian filter is required to classify experimental traces, training the network with 10$^{6}$ experimental traces is of no advantage. On the synthetic data, we find a network trained with synthetic traces to yield a similar error rate than the Bayesian filter, while it slightly outperforms the latter in terms of robustness versus variations in experimental parameters such as SNR, the signal current offset and the tunnel couplings. This advantage is even more pronounced for the classification of real measurement data: our neural network trained with a combination of synthetic data and measured noise outperforms the Bayesian benchmark by 7 \%, as seen from the visibility of Rabi oscillations of the spin qubit. Here, the combination of an absence of labelling errors in the training data and the setup-specific noise proves to be particularly advantageous.

Given that that our results should be representative for qubit types with stochastic readout schemes 
and that the realtime performance of the neural network can be further optimized by running it on dedicated hardware, neural networks can represent an important building block for cryoelectronics yielding high-fidelity readout in scalable qubit architectures.

\section{Acknowledgements}
We thank Uwe Klemradt and Hendrik Bluhm for valuable discussion. This work has been funded by the German Research Foundation (DFG) within the projects BO 3140/4-1, 289786932 and the cluster of excellence "Matter and light for quantum computing" (ML4Q) as well as by the Federal Ministry of Education and Research under Contract No. FKZ: 13N14778. Project Si-QuBus received funding from the QuantERA ERA-NET Cofund in Quantum Technologies implemented within the European Union's Horizon 2020 Programme.

\section{Author contributions}
T.S. and J.L. developed the network architecture. T.S. did the computation and analysis of the data assisted by F.S. and A.S. and supported by L.R.S. T.S. and A.H. measured noise spectra and experimental signal traces on a sample fabricated by F.S., A.S. and D.B. L.R.S. conceived and supervised the study and all authors discussed the results. T.S., A.S., D.B. and L.R.S. wrote the manuscript, which all other authors reviewed.

\section{Competing interests}
The authors declare no competing interests.

\section{Data availability}

The datasets generated during and/or analysed during the current study are available from the corresponding author on reasonable request.


\end{document}